\documentclass[mpicover]{mpireport-inf}
\usepackage{times}
\usepackage{url}
\usepackage{amsmath}
\usepackage{stmaryrd}
\usepackage{defs}
\usepackage{verbatim}
\usepackage{pstricks,pst-node}
\usepackage{graphicx}
\usepackage{cite}
\usepackage{amssymb}

\newgray{lgray}{.7}

\title{On Verifying Complex Properties using Symbolic Shape Analysis}

\author{Thomas Wies \and Viktor Kuncak \and \\ Karen Zee \and Andreas
  Podelski \and \\ Martin Rinard}

\begin{addresses}
Thomas Wies\\
Max-Planck-Institut f\"ur Informatik\\
Saarbr\"ucken, Germany \bigskip\\
Viktor Kuncak\\
MIT Computer Science and Artificial Intelligence Lab\\
Cambridge, USA \bigskip\\
Karen Zee\\
MIT Computer Science and Artificial Intelligence Lab\\
Cambridge, USA \bigskip\\
Andreas Podelski\\
Max-Planck-Institut f\"ur Informatik\\
Saarbr\"ucken, Germany \bigskip\\
Martin Rinard\\
MIT Computer Science and Artificial Intelligence Lab\\
Cambridge, USA\\
\end{addresses}

\repnumber{2--001}                 

\setyear{2006}
\setrepyear{2006}
\setmonth{April}

\begin{document}

\begin{abstract}
  One of the main challenges in the verification of software systems
  is the analysis of unbounded data structures with dynamic memory
  allocation, such as linked data structures and arrays.  We describe
  Bohne, a new analysis for verifying data structures.  Bohne verifies
  data structure operations and shows that 1) the operations preserve
  data structure invariants and 2) the operations satisfy their
  specifications expressed in terms of changes to the set of objects
  stored in the data structure.  During the analysis, Bohne infers
  loop invariants in the form of disjunctions of universally
  quantified Boolean combinations of formulas, represented as sets of
  binary decision diagrams.  To synthesize loop invariants of this
  form, Bohne uses a combination of decision procedures for Monadic
  Second-Order Logic over trees, SMT-LIB decision procedures
  (currently CVC Lite), and an automated reasoner within the Isabelle
  interactive theorem prover.  This architecture shows that
  synthesized loop invariants can serve as a useful communication
  mechanism between different decision procedures.  In addition, Bohne
  uses field constraint analysis, a combination mechanism that enables
  the use of uninterpreted function symbols within formulas of Monadic
  Second-Order Logic over trees.  Using Bohne, we have verified
  operations on data structures such as linked lists with iterators
  and back pointers, trees with and without parent pointers, two-level
  skip lists, array data structures, and sorted lists.  We have
  deployed Bohne in the Hob and Jahob data structure analysis systems,
  enabling us to combine Bohne with analyses of data structure clients
  and apply it in the context of larger programs.  This report
  describes the Bohne algorithm as well as techniques that Bohne uses
  to reduce the ammount of annotations and the running time of the analysis.
\end{abstract}

\makempicover

\tableofcontents

\chapter{Introduction}

Complex data structure invariants are one of the main challenges in
verifying software systems.  Unbounded data
structures such as linked data structures and dynamically
allocated arrays make the state space of software artifacts
infinite and require new reasoning techniques (such as
reasoning about reachability) that have traditionally not
been part of theorem provers specialized for program
verification.  The ability of linked structures to 
change their shape makes them a powerful programming
construct, but at the same time makes them difficult to
analyze, because the appropriate analysis representation is
dependent on the invariants that the program maintains.  It is therefore
not surprising that the most successful
verification approaches for analysis of data structures use
parameterized abstract domains; these analyses include
parametric shape analysis \cite{SagivETAL02Parametric} as
well as predicate abstraction
\cite{BallETAL01AutomaticPredicateAbstraction,
HenzingerETAL02LazyAbstraction} and its generalizations
\cite{FlanaganQadeer02PredicateAbstraction,
LahiriBryant04IndexedPredicateDiscoveryUnboundedSystemVerification}.

This paper presents \emph{Bohne}, an algorithm for inferring
loop invariants of programs that manipulate heap-allocated
data structures.  Like predicate abstraction, Bohne is
parameterized by the properties to be verified.  What makes
the Bohne algorithm unique is the use of a precise
abstraction domain that can express detailed properties of
different regions of programs infinite memory, and a range
of techniques for exploring this analysis domain using
decision procedures.  The algorithm was initially developed
as a symbolic shape analysis
\cite{Wies04SymbolicShapeAnalysis,
PodelskiWies05BooleanHeaps} for linked data structures and
uses the key idea of shape analysis: the partitioning of
objects according to certain unary predicates.  One of the
observations of our paper is that the synthesis of heap
partitions is not only useful for analyzing shape
properties (which involve transitive closure), but also for combining such shape properties
with sorting properties
of data structures and properties expressible using linear
arithmetic and first-order logic.  

We next put the core Bohne algorithm in the context of
predicate abstraction and parametric shape analysis
approaches.

\smartparagraph{Predicate abstraction.}  Bohne builds on predicate
abstraction but introduces important new techniques that make it
applicable to the domain of shape analysis.  There are two main
sources of complexity of loop invariants in shape analysis.  The first
source of complexity is the fact that the invariants contain
reachability predicates.  To address this problem, Bohne uses a
decision procedure for monadic second-order logic over trees
\cite{KlarlundETAL00MONA}, and combines it with uninterpreted function
symbols in a way that preserves completeness in important cases
\cite{WiesETAL06FieldConstraintAnalysis}.  The second source of
complexity is that the invariants contain universal quantifiers in an
essential way.  Among the main approaches for dealing with quantified
invariants in predicate abstraction is the use of Skolem constants
\cite{FlanaganQadeer02PredicateAbstraction}, indexed predicates
\cite{LahiriBryant04IndexedPredicateDiscoveryUnboundedSystemVerification}
and the use of abstraction predicates that contain quantifiers.  The
key difficulty in using Skolem constants for shape analysis is that
the properties of individual objects depend on the ``context'', given
by the properties of surrounding objects, which means that it is not
enough to use a fixed Skolem constant throughout the analysis, it is
instead necessary to instantiate universal quantifiers from previous
loop iterations, in some cases multiple times.  Compared to indexed
predicates
\cite{LahiriBryant04IndexedPredicateDiscoveryUnboundedSystemVerification}
the domain used by Bohne is more general because it contains
disjunctions of universally quantified statements.  The presence of
disjunctions is not only more expressive in principle, but allows
Bohne to keep formulas under the universal quantifiers more specific.
This enables the use of less precise, but more efficient algorithms
for computing changes to properties of objects without 
losing too much precision in the overall analysis.  Finally, the
advantage of using abstraction tailored to shape analysis compared to
using quantified global predicates is that the parameters to
shape-analysis-oriented abstraction are properties of objects in a
state, as opposed to global properties of a state, and the number of
global predicates needed to emulate state predicates is exponential in
the number of properties
\cite{ManevichETAL05PredicateAbstractionCanonicalAbstractionSinglyLinkedLists,
  Wies04SymbolicShapeAnalysis}.

\smartparagraph{Shape analysis.}
Shape analyses are precise
analyses for linked data structures.  They were originally
used for compiler optimizations
\cite{MuchnikJones81ProgramFlowAnalysis, GhiyaHendren95ConnectionAnalysis,
GhiyaHendren96TreeOrDag} and lacked
precision needed to establish invariants that Bohne is analyzing.
Precise data structure analysis for the purpose of verification include
\cite{KuncakETAL02RoleAnalysis,
FradetMetayer97ShapeTypes, KlarlundSchwartzbach93GraphTypes,
LeeETAL05AutomaticVerificationPointerProgramsUsing,
Moeller01PALE, SagivETAL02Parametric} and have recently also been applied to
verify set implementations \cite{Reineke05ShapeAnalysisSets}.
Unlike Bohne, most shape analyses
that synthesize loop invariants are based on precomputed transfer
functions and a fixed (though parameterized) set of properties to be tracked; recent approaches
enable automation of such computation using decision procedures 
\cite{YorshETAL05AutomaticAssumeGuaranteeReasoningHeap,
YorshETAL04SymbolicallyComputingMostPrecise,
YorshETAL05LogicalCharacterizationsHeapAbstractions,
PodelskiWies05BooleanHeaps, WiesETAL06FieldConstraintAnalysis}
or finite differencing
\cite{RepsETAL03FiniteDifferencingLogicalFormulasStaticAnalysis}.  We are
currently working on an effort to compare such
different analysis on a joint set of benchmarks
\cite{KuncakETAL06ProposalEstablishShapeAnalysisBenchmarks}.
Our approach differs from
\cite{LahiriQadeer06VerifyingPropertiesWellFoundedLinkedLists}
in using complete reasoning about reachability in both lists
and trees, and using a different architecture of the
reasoning procedure.  Our reasoning procedure uses a
coarse-grain combination of reachability reasoning with
decision procedures and theorem provers for numerical and
first-order properties, as opposed to using a Nelson-Oppen
style theorem prover.  This allowed us to easily combine
several tools that were developed completely independently
\cite{KlarlundETAL00MONA, BarrettBerezin04CVCLite,
NipkowETAL02IsabelleHOL}.  Shape analysis approaches have
also been used to verify sortedness properties
\cite{LevAmiETAL00PuttingStaticAnalysisWorkVerification} relying 
on manually abstracting sortedness relation.

Recently there has been a resurgence of decision procedures and
analyses for linked list data structures
\cite{BalabanETAL05ShapeAnalysisPredicateAbstraction,
  DistefanoETAL06LocalShapeAnalysisBasedSeparationLogic,
  ManevichETAL05PredicateAbstractionCanonicalAbstractionSinglyLinkedLists,
  BinghamRakamaric05LogicDecisionProcedure,
  RaniseZarba05DecidableLogicPointerProgramsManipulatingLinkedLists},
where the emphasis is on predictability (decision procedures for
well-defined classes of properties of linked lists), efficiency
(membership in NP), the ability to interoperate with other reasoning
procedures, and modularity.  Although the Bohne approach is not
limited to lists, it can take advantage of decision procedures for
lists by applying such specialized decision procedures when they are
applicable and using more general reasoning otherwise.

Bohne could also take advantage of logics for reasoning about
reachability, such as the logic of reachable shapes
\cite{YorshETAL06LogicReachablePatternsLinkedDataStructures}.
Existing logics, such as guarded fixpoint logic
\cite{Graedel99DecisionProceduresGuardedLogics} and description logics
with reachability
\cite{CalvaneseETAL99ReasoningExpressiveDescriptionLogicsFixpoints,GeorgievaMaier05DescriptionLogicsForShapeAnalysis}
are attractive because of their expressive power, but so far no
decision procedures for these logics have been implemented.  Automated
theorem provers such as Vampire \cite{Voronkov95Vampire} and SPASS
\cite{Spass} can be used to reason about properties of linked data
structures, but axiomatizing reachability in first-order logic is
non-trivial in practice
\cite{Nelson83VerifyingReachabilityInvariantsLinkedStructures,
  LevAmiETAL05SimulatingReachabilityFirstOrderLogic} and not possible
in general.

\section{Contributions}

We have previously described the general idea of symbolic
shape analysis
\cite{PodelskiWies05BooleanHeaps} as well as the field constraint analysis
decision procedure for combining reachability reasoning with 
uninterpreted function symbols
\cite{WiesETAL06FieldConstraintAnalysis}.   
In \cite{ZeeETAL04CombiningTheoremStaticAnalysisDataStructureConsistency}
we have described splitting of proof obligations in the
context of verifying proof obligations using the Isabelle
interactive theorem prover.  One of the insights in this
paper is that such splitting can be an effective way of
combining different reasoning procedures during fixpoint
computation in abstract interpretation.  These previous
techniques are therefore the starting point of this paper.
The main contributions of this paper are the following:
\begin{enumerate}

\item We present a technique for combining different decision
  procedures through 1) a static analysis that synthesizes Boolean
  algebra expressions over sets defined by arbitrary abstraction
  predicates, 2) a proof obligation splitting approach that discharges
  different conjuncts using different decision procedures, and 3) a
  verification-condition generator that preserves abstract variables.
  This approach addresses a key question in extending a Nelson-Oppen
  style combination to theories that share \emph{sets of elements}.
  In general, such combination would require guessing and propagating
  an exponential number of Boolean algebra expressions.  In our
  approach, symbolic shape analysis \cite{PodelskiWies05BooleanHeaps}
  synthesizes Boolean algebra expressions that are used as assumptions
  in decision procedures calls and are therefore shared by all
  participating decision procedures.

\item We describe a method for synthesis of Boolean heap programs that
  improves the efficiency of fixpoint evaluation by precomputing
  abstract transition relations and can control the
  precision/efficiency trade-off by recomputing transition relations
  on-demand during fixpoint computation.

\item We introduce semantic caching of decision procedure queries
  across different fixpoint iterations and even different analyzed
  procedures.  The caching yields substantial improvements for
  procedures that exhibit some similarity, which opens up the
  possibility of using our analysis in an interactive context.


\item We describe a static analysis that propagates
      precondition conjuncts and quickly finds many true
      facts, reducing the running time and the
      number of needed abstraction predicates for the
      subsequent symbolic shape analysis.

\item We present a domain-specific quantifier instantiation 
      technique that often eliminates the need for the
      underlying decision procedures to deal with quantifiers.

\end{enumerate}
Together, these new techniques allowed us to verify a range
of data structures without specifying loop invariants and
without specifying a large number of abstraction predicates.
Our examples include implementations of lists (with
iterators and with back pointers), trees with parent
pointers, and sorted lists.  What makes these results
particularly interesting is a higher level of automation
than in previous approaches: Bohne synthesizes loop
invariants that involve reachability expressions and
numerical quantities, yet it does not have precomputed
transfer functions for a particular set of abstraction
predicates.  Bohne instead uses decision procedures to
reason about arbitrary predicates definable in a given
logic.  Moreover, in our system the developer is not
required to manually specify the changes of membership of
elements in sets because such changes are computed by Bohne
and used to communicate information between different
decision procedures.


\smartparagraph{Bohne as component of Hob and Jahob.}  Bohne is part
of the data structure verification frameworks Hob
\cite{LamETAL04HobProjectWebPage,LamETAL05HobTool} and Jahob
\cite{Kuncak06JahobProjectWebPage}. The goal of these systems is to
verify data structure consistency properties in the context of
non-trivial programs. To achieve this goal, these tools combine
multiple static analyses, theorem proving, and decision procedures.
In this paper we present our experience in deploying Bohne in the
Jahob framework.  The input language for Jahob is a subset of Java
extended with annotations written as special comments. Therefore,
Jahob programs can be compiled and executed using existing Java
compilers and virtual machines.

\begin{figure}
  \begin{center}
  \vspace*{-1em}
  \includegraphics[scale=0.4]{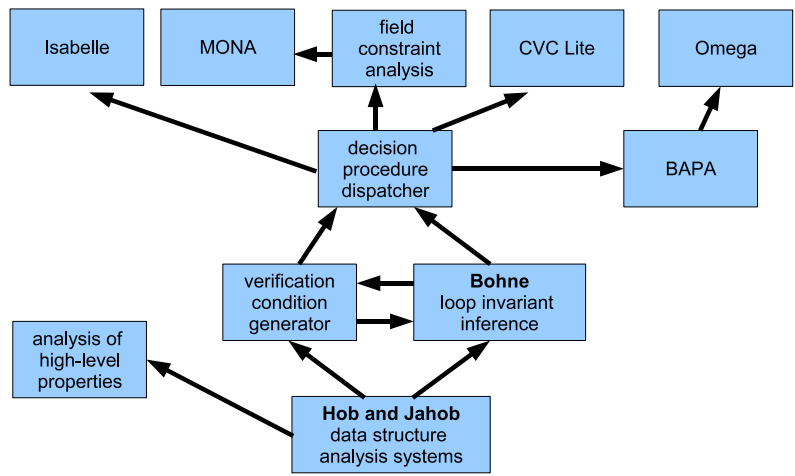}
  \vspace*{-0.5in}
  \end{center}
  \caption{Architecture of the Hob and Jahob Data Structure Analysis Systems}  
  \label{fig:JahobArchitecture}
\end{figure}

Figure \ref{fig:JahobArchitecture} illustrates the
integration of Bohne into the Jahob framework. 
Bohne uses Jahob's
facilities for symbolic execution of program statements and
the validity checker to compute the abstraction
of the source program.
The output of
Bohne is the source program annotated with the inferred loop
invariants. The annotated program serves as an input to a
verification condition generator. The generated verification
conditions are verified using a validity checker that
combines special purpose decision procedures, a general
purpose theorem prover, and reasoning techniques such as
field constraint analysis
\cite{WiesETAL06FieldConstraintAnalysis}. 

\chapter{Motivating Example}

We illustrate our technique on the procedure \q{SortedList.insert} shown
in Figure~\ref{fig:SortedListExample}. This procedure inserts a \q{Node}
object into a global sorted list. The annotation given by special
comments \q{/*: \ldots\ */} consists of data structure invariants, pre- and
postconditions, as well as hints for the analysis. Formulas are
expressed in a subset of the language used in the Isabelle interactive
theorem prover \cite{NipkowETAL02IsabelleHOL}. The specification uses
an abstract set variable \q{content} which is defined as the set of
non-null objects reachable from the global variable \q{first} by
following field \q{Node.next}.  The construct \verb+rtrancl_pt+ is a
higher-order function that maps a binary predicate to its reflexive
transitive closure. The data structure invariants are specified by the
annotation \q{invariant "\ldots"}. For instance, the first invariant
expresses the fact that the field \q{Node.next} forms trees in the
heap, i.e. that \q{Node.next} is acyclic and injective; the third
invariant expresses the fact that the elements stored in the list are
sorted in increasing order according to field \q{Node.data}.  The
precondition of the procedure, \q{requires "\ldots"}, states that the object
to be inserted is non-null and not yet contained in the list. The
postcondition, \q{ensures "\ldots"}, expresses that the content of the list
is unchanged except for the argument being added.

\begin{figure}
  \footnotesize
  \verbatiminput{Node.java}
  \verbatiminput{SortedList.java}
  \normalsize
  \caption{\label{fig:SortedListExample}Insertion into a sorted list}
\end{figure}

The loop in the procedure body traverses the list until it
finds the proper position for insertion. It then inserts the
argument such that the resulting data structure is again a
sorted list.  Our analysis, Bohne, is capable of verifying
that the postcondition holds at the end of the procedure
\q{insert}, that data structure invariants are preserved,
and that there are no run-time errors such as null pointer
dereferences.  In order to establish these properties, Bohne
derives a complex loop invariant shown in
Fig.~\ref{fig:SortedListInvariant}.
\begin{figure}
\footnotesize
\begin{center}
\begin{verbatim}
tree [Node.next] &
(first = null | (ALL n. n..Node.next ~= first)) &
(ALL v. v : content & v..Node.next ~= null --> 
  v..Node.data <= v..Node.next..Node.data) &
(ALL v w. v ~= null & w ~= null & v..Node.next = w --> 
  w : content) &
n ~= null & n ~: content &
reach_curr = {v. rtrancl_pt (% x y. x..Node.next = y) curr v} &
content = old content &
(curr ~= null --> curr : content) & 
(prev = null --> first = curr) &
(prev ~= null --> 
  prev : content & prev ~: reach_curr & prev..Node.next = curr) &
(ALL v. v ~: reach_curr & v : content --> v : lt_n)
\end{verbatim}
\end{center}
\caption{\label{fig:SortedListInvariant} Loop invariant for procedure \q{SortedList.insert}}
\end{figure}
The main difficulties for inferring this invariant are: (1) it
contains universal quantifiers over an unbounded domain and (2) it
requires reasoning over multiple theories, here reasoning over
reachability, reasoning over numerical domains, and reasoning over
uninterpreted function symbols.

Bohne infers universally quantified invariants using symbolic shape
analysis based on Boolean heaps \cite{Wies04SymbolicShapeAnalysis,
  PodelskiWies05BooleanHeaps}. This approach can be viewed as a generalization of
predicate abstraction or a symbolic approach to parameteric shape analysis.
 Abstraction predicates can be Boolean-valued
state predicates (which are either true or false in a given state,
such as \verb+curr_prev+) or predicates denoting sets of heap objects in
a given state (which are true of a \emph{given object} in a
\emph{given state}, such as \verb+lt_n+). The latter serve as building
blocks of the inferred universally quantified invariants. The
\q{track(\ldots)} annotation is used as a hint on which predicates the
analysis should use for the abstraction of which code fragments.

To reduce the annotation burden we use a syntactic analysis to infer
abstraction predicates automatically (e.g. predicate \verb+reach_curr+
in the loop invariant). Furthermore, parts of the invariant often
literally come from the procedure's precondition. In particular, data
structure invariants are often preserved as long as the heap is not
mutated. We therefore precede the symbolic shape analysis phase with
an analysis that propagates precondition conjuncts accross the
control-flow graph of the procedure's body.  Using this propagation
technique we are able to infer the first six conjuncts of the
invariant. The symbolic shape analysis phase makes use of this partial
invariant to infer the full invariant shown in
Fig.~\ref{fig:SortedListInvariant}.

Bohne's symbolic shape analysis enables the combination of decision
procedures by connecting the analysis with a proof obligation
splitting approach that discharges different conjuncts using different
decision procedures, and a verification-condition generator that
preserves abstract variables.  Thereby the inferred invariants
communicate information between different decision procedures.  This
combination is best illustrated with an example.
Figure~\ref{fig:SortedListVC} shows one of the generated verification
conditions for the procedure \q{SortedList.insert}. It expresses the
fact that the sortedness property is reestablished after executing the
path from the exit point of the loop through the if-branch of the
conditional to the procedure's return point. The symbol ``\q{I}'' denotes the
loop invariant given in Fig.~\ref{fig:SortedListInvariant}. This
verification condition is valid. Its proof requires the fact
\[\q{content'} = \q{content} \; \q{Un} \; \set{\q{n}}\]
Denote this fact $P$.  $P$ follows from the given
assumptions. The MONA decision procedure is able to conclude $P$
by expanding the definitions of the abstract sets
\q{content} and \q{content'}.  However, MONA is not able to prove the
verification condition, because proving its conclusion requires
reasoning over integers. On the other hand, the CVC Lite decision
procedure is able to prove the conclusion given the fact $P$ by
reasoning over the abstract sets without expanding their definitions, but
is not able to conclude $P$ from the assumptions,
because this deduction step requires reasoning over reachability. In
order to communicate $P$ between the two decision procedures, symbolic
shape analysis infers, in addition to the loop invariant \q{I}, an invariant
for the procedure's return point that includes the missing fact $P$.
This invariant enables CVC Lite to prove the verification
condition.

\begin{figure}
  \begin{center}
    \footnotesize
\begin{verbatim}
I & ~(curr..Node.data < n..Node.data) & prev ~= null &
Node.next' = Node.next[n := curr][prev := n] &
content' = 
  {v. v ~= null & rtrancl_pt (% x y. x..Node.next' = y) first v} &
v : content' & n..Node.next' ~= null --> 
  v..Node.data <= v..Node.next'..Node.data
\end{verbatim}
  \end{center}
  \caption{\label{fig:SortedListVC} Verification condition for
    preservation of sortedness}
\end{figure}

\chapter{The Bohne Algorithm}

We next describe the symbolic shape analysis algorithm implemented in
Bohne. What makes this algorithm unique is the fact that abstract
transition relations are computed on-demand in each fixpoint iteration
taking into account the \emph{context} in form of already explored
abstract states. 
This approach allows the algorithm to take advantage of precomputed
abstract transition relations from previous fixpoint iterations, 
while maintaining sufficient precision for the analysis of linked
data structures by recomputing the transitions when the context changes
in a significant way.

\section{Reachability Analysis}

The input of Bohne is the procedure to be analyzed,
preconditions specifying the initial states of the procedure, and a
set of abstraction predicates. Bohne converts the procedure into a set
of guarded commands that correspond to the loop-free paths in the
control-flow graph.

\begin{figure}
  \begin{equation*}
    \begin{array}{l}
      \kw{proc}\ \m{Reach}(
      \begin{array}[t]{@{}l}
        \init : \mbox{precondition formula}, \\
        \ell_{\init} : \mbox{initial program location}, \\ 
        T :  \mbox{set of guarded commands}) =
      \end{array}\\
      \quad 
      \begin{array}{l}
        \kw{let}\ \init^\# = \m{abstract}(\init) \\
        \kw{let}\ \m{root} = \langle \m{location} = \ell_\init; \m{states} =
        \init^\#; \m{sons} = \emptyset \rangle \\
        \kw{let}\ \m{unprocessed} = \set{\m{root}} \\
        \kw{while}\ \m{unprocessed} \neq \emptyset \ \kw{do} \\
        \quad
        \begin{array}{l}
          \kw{choose}\ n \in \m{unprocessed} \\
          \kw{for all}\ (n.\m{location}, c, \ell') \in T \ \kw{do} \\
          \quad 
          \begin{array}{l}
            \kw{let}\ \m{context} = \pset{m.\m{states}}{m.\m{location} = \ell}\\
            \kw{let}\ \m{old} = \pset{m.\m{states}}{m.\m{location} = \ell'} \\
            \kw{let}\ \m{new} =
            \m{AbstractPost}(c,\m{context},n.\m{states}) - \m{old}\\
            \kw{if}\ \m{new} \neq \emptyset \ \kw{then}\ \\
            \quad
            \begin{array}{l}
              \kw{let}\ n' = \langle \m{location} = \ell'; \m{states} =
              \m{new}; \m{sons} = \emptyset\rangle \\ 
              n.\m{sons} := n.\m{sons} \cup \set{(c,n')} \\
              \m{unprocessed} := \m{unprocessed} \cup \set{n'}
            \end{array} \\
          \end{array} \\
          \m{unprocessed} := \m{unprocessed} - \set{n}
        \end{array} \\
        \kw{return}\ \m{root}
      \end{array}
    \end{array}
  \end{equation*}
  \caption{Reachability analysis in Bohne}
  \label{fig:Algorithm}
\end{figure}

Figure \ref{fig:Algorithm} gives the pseudo code of Bohne's top-level
fixpoint computation loop. The analysis first abstracts the
conjunction of the procedure's preconditions obtaining an initial set
of abstract states.  It then computes an abstract reachability tree in
the spirit of lazy abstraction \cite{HenzingerETAL02LazyAbstraction}.
Each node in this tree is labeled by a program location and a set of
abstract states, the root being labeled by the initial location and
the abstraction of the preconditions. The edges in the tree are
labeled by guarded commands. The reachability tree keeps track of
abstract traces which are used for the analysis of abstract
counterexamples.

For each unprocessed node in the tree, the analysis computes the
abstract postcondition for the associated abstract states and all
outgoing transitions of the corresponding program location.
Transitions are abstracted on-demand taking into account the already
discovered reachable abstract states for the associated program
location.  Whenever the difference between the already discovered
abstract states of the post location and the abstract post states of
the processed transition is non-empty, a new unprocessed node is added
to the tree.  The analysis stops after the list of unprocessed nodes
becomes empty, indicating that the fixpoint is reached. After
termination of the reachability analysis, Bohne annotates the original
procedure with the computed loop invariants and passes the result to
the verification condition generator, which verifies that the inferred
loop invariants are sufficient to prove the target properties.

The algorithm in Figure \ref{fig:Algorithm} is parameterized by the
abstract domain and its associated operators. An abstract state of the
analysis is given by a set of bitvectors over abstraction predicates
which we call a Boolean heap. It corresponds to a universally
quantified Boolean combination of abstraction predicates.  A Boolean
heap describes all concrete states whose heap is partitioned according
to the bitvectors in the Boolean heap.  Focusing on algorithmic
details, we now give a detailed description of the abstract domain,
abstraction function, and the abstract post operator.


\section{Symbolic Shape Analysis}

Following the framework of abstract interpretation
\cite{CousotCousot77AbstractInterpretation}, a static analysis is
defined by lattice-theoretic domains and by fixpoint iteration over
the domains. Symbolic shape analysis can be seen as a generalization
of predicate abstraction
\cite{GrafSaidi97ConstructionAbstractStateGraphsPVS}.  For
\emph{predicate abstraction} the analysis computes an invariant; the
fixpoint operator is an abstraction of the \emph{post} operator; the
concrete domain consists of sets of states (represented by closed
formulas), and the abstract domain of a finite lattice of closed
formulas.

\smartparagraph{Abstract Domain.} Let $\Pred$ be a finite set of
abstraction predicates $p(\freevar)$ with an implicit free variable
$\freevar$ ranging over heap objects. A \emph{cube} $\cube$ is a
partial mapping from $\Pred$ to $\set{0,1}$. We call a total cube
\emph{complete}. We say that predicate $p$ occurs positively (occurs
negatively, does not occur) in $\cube$ if $\cube(p) = 1$ ($\cube(p) =
0$, $\cube(p)$ is undefined). We denote by $\Cubes$ the set of all
cubes. An abstract state is a subset of cubes, which we call a
\emph{Boolean heap}. The abstract domain is given by sets of Boolean
heaps, i.e. sets of sets of cubes:
\[\AbsDom = 2^{2^{\Cubes}}\enspace.\] 
\smartparagraph{Meaning Function.} The meaning function $\gamma$ is defined
on cubes, Boolean heaps, and sets of Boolean heaps as follows:
\[\gamma (\cube) = \Conj_{p \in \Pred \cap \dom(C)} p^{C(p)}, 
\qquad 
\gamma (\absstate) = \All{\freevar}{\Disj_{\cube \in \absstate} \gamma(\cube)},
\qquad
\gamma(\absstates) = \Disj_{\absstate \in \absstates} \gamma(\absstate)\enspace.
\]
The meaning of a cube $\cube$ is the conjunction of the predicates in
$\Pred$ and their negations. A concrete state is represented by a
Boolean heap $\absstate$ if all objects in the heap are represented by
some cube in $\absstate$. The meaning of a set $\absstates$ of Boolean
heaps is the disjunction of the meaning of all its elements.

\smartparagraph{Lattice Structure.}  Define a partial order $\sqsubseteq$ on
cubes by:
$$\cube \sqsubseteq \cube' \Def{\iff} \all{p \in \Pred}{
  \cube'(p) = \cube(p) \ \lor\ (\cube'(p) \text{ is undefined}})
\enspace.$$ For a cube $\cube$ and Boolean heap $\absstate$ we write
$\cube \in_c \absstate$ as a short notation for the fact that $\cube$ is
complete and there exists $\cube' \in \absstate$ such that $\cube \sqsubseteq
\cube'$. The partial order $\sqsubseteq$ is extended from cubes to a preorder on
Boolean heaps:
\begin{align*}
  \absstate \sqsubseteq \absstate' & \Def{\iff} \all{\cube \in
    \absstate}{\ex{\cube' \in \absstate'}{\cube \sqsubseteq \cube'}} \enspace.
\end{align*}
For notational convenience we identify Boolean heaps up to subsumption
of cubes, i.e. up to equivalence under the relation ($\sqsubseteq \cap \sqsubseteq^{-1}$).
We then identify $\sqsubseteq$ with the partial order on the corresponding
quotient of Boolean heaps. In the same way we extend $\sqsubseteq$ from Boolean
heaps to a partial order on the abstract domain. These partial orders
induce Boolean algebra structures. We denote by $\sqcap$, $\sqcup$ and $\compl{\
  \cdot \ }$ the meet, join and complement operations of these Boolean
algebras.  Boolean heaps, the abstract domain, and operations of the
Boolean algebras are implemented using BDDs
\cite{Bryant86GraphBasedAlgorithmsBooleanFunctionManipulation}.

\smartparagraph{Context-sensitive Cartesian post.} The abstract post
operator implemented in Bohne is a refinement of the abstract post
operator on Boolean heaps that is presented in
\cite{PodelskiWies05BooleanHeaps}. Its core is given by the
\emph{context-sensitive Cartesian post operator}.  This operator maps
a guarded command $c$, a formula $\Gamma$, and a set of Boolean heaps
$\absstates$ to a set of Boolean heaps as follows:
\[
\begin{array}{l}
  \m{CartesianPost} (c, \context, \absstates) = \\
\quad \pset{\pset{\bigsqcap \pset{\cube'}{\all{p \in \Pred}{\cube
      \sqsubseteq \wlp^\#(c, \context, p^{\cube'(p)})}}}{\cube \in_c \absstate}}{\absstate \in \absstates}. 
\end{array}
\] 
The actual abstraction is hidden in the computation of the function
$\wlp^\#$ which is defined by:
\[\wlp^\# (c , \Gamma, F) = \pset{\cube}{\Gamma \conj \gamma (\cube) \models \wlp (c,
  F)} \enspace.\] 
The Cartesian post maps each Boolean heap $\absstate$ in $\absstates$
to a new Boolean heap $\absstate'$. For a given state $s$ satisfying
$\gamma(\absstate)$, a cube $\cube$ in $\absstate$ represents a set of heap
objects in $s$. The Cartesian post computes the local effect of
command $c$ on each set of objects which is represented by some
complete cube in $\absstate$: each complete cube $\cube$ in
$\absstate$ is mapped to the smallest cube $\cube'$ that represents at
least the same set of objects in the post states under command $c$.
Consequently each object in a given post state is represented by some
cube in the resulting Boolean heap $\absstate'$, i.e. all post states
satisfy $\gamma(\absstate')$.  The effect of $c$ on the objects represented
by some cube is expressed in terms of weakest preconditions of
abstraction predicates. These are abstracted by the function $\wlp^\#$.

Computing the effect of $c$ for each cube in $\absstate$ locally
implies that we do not take into account the full information provided
by $\absstate$. In principle one can strengthen the abstraction of
weakest preconditions by taking into account the Boolean heap for
which the post is computed: $\wlp^\#(c, \gamma(\absstate), p)$.  The
abstract post would be more precise, but as a consequence abstract
weakest preconditions would have to be recomputed for each Boolean
heap. This would make the analysis infeasible.  Nevertheless, such
global context information is valuable when updated predicates
describe global properties such as reachability. Therefore, we would
like to strengthen the abstraction using some global information,
accepting that abstract weakest preconditions have to be recomputed
occasionally.  The formula $\context$ allows this kind of
strengthening. It is the key tuning parameter of the analysis. We
impose a restriction on $\context$ to ensure soundness: we say that
$\context$ is a \emph{context formula} for a set of Boolean heaps
$\absstates$ if $\gamma(\absstates)$ implies $\context$.  Restricting the
Cartesian post to context formulas ensures soundness with respect to
the best abstract post operator on sets of Boolean heaps.

\begin{figure}
  \begin{equation*}
    \begin{array}{l}
      \begin{array}{rcl}
        \kw{proc}\ \m{CartesianPost}(
        \begin{array}[t]{@{}l}
          c : \mbox{guarded command}, \\ 
          \context : \mbox{context formula}, \\ 
          \absstates : \AbsDom) : \AbsDom =
        \end{array}
      \end{array}\\
      \quad
      \begin{array}{l}
        \kw{let}\ \m{c^\#} = \Cubes\\
        \kw{if}\ c^\#\  \mbox{is precomputed for}\ 
        (c,\context)\ \kw{then}\ 
        c^\# := \mbox{lookup}(c,\Gamma)\\
        \kw{else}\ 
        \kw{foreach}\ p \in \Pred \ \kw{do}\\
          \quad 
          \begin{array}{l}
            c^\# := c^\# \sqcap \left(
              \begin{array}{l}
                [p' \mapsto 1] \sqcap \compl{\wlp^\#(c, \context, \lnot p)} \; \sqcup \\
                \mbox{} [p' \mapsto 0] \sqcap \compl{\wlp^\#(c, \context,p)} 
              \end{array} \right)
          \end{array}\\
        \kw{let}\ \absstates' = \emptyset\\
        \kw{foreach}\ \absstate \in \absstates \ \kw{do}\\
        \quad 
        \begin{array}{l}
          \kw{let}\ \absstate' = \m{RelationalProduct}(\absstate, c^\#)\\
          \absstates' := \absstates' \sqcup \set{\absstate'}
        \end{array}\\
        \kw{return}\ \absstates'
      \end{array}
    \end{array}
  \end{equation*}
  \caption{Context-sensitive Cartesian post}
  \label{fig:CartesianPost}
\end{figure}

Figure~\ref{fig:CartesianPost} gives an implementation of the
Cartesian post operator that exploits the representation of Boolean
heaps as BDDs. First it precomputes an abstract transition relation
$c^\#$ which is expressed in terms of cubes over primed and unprimed
abstraction predicates. After that it computes the relational product
of $c^\#$ and each Boolean heap. The relational product conjoins a
Boolean heap with the abstract transition relation, projects the
unprimed predicates, and renames primed to unprimed predicates in the
resulting Boolean heap. Note that that the abstract transition
relation only depends on the abstracted command $c$ and the context
formula $\context$. This allows us to cache abstract transition
relations and avoid their recomputation in later fixpoint iterations
where $\context$ is unchanged.

\smartparagraph{Splitting.} The Cartesian post operator maps each
Boolean heap in a set of Boolean heaps to one Boolean heap. This means
that in terms of precision the Cartesian post does not exploit the
fact that the abstract domain is given by \emph{sets} of Boolean
heaps. In the following we describe an operation that splits a Boolean
heap into a set of Boolean heaps. The splitting maintains important
invariants of Boolean heaps that result from best abstractions of
concrete states. We split Boolean heaps before applying the Cartesian
post. This increases the precision of the analysis by carefully
exploiting the disjunctive completeness of the abstract domain.

Traditional shape analysis uses the idea of summary nodes to
distinguish abstract objects that represent multiple concrete objects
from abstract objects that represent single objects. This information
is useful for increasing the precision of the abstract post operator.
We can mimic this idea by adding abstraction predicates that denote
singleton sets, e.g. by adding predicates expressing properties such
as that an object is pointed to by some local variable. If a Boolean
heap $\absstate$ is the best abstraction of some concrete state then
for every \emph{singleton predicate} $p$ it contains exactly one
complete cube with a positive occurrence of $p$. Boolean heaps
resulting from the Cartesian post typically do not have this property
which makes the analysis imprecise. Therefore we split each Boolean
heap before application of the Cartesian post into a set of Boolean
heaps such that the above property is reestablished. Let $P$ be the
subset of abstraction predicates denoting singletons then the
\emph{splitting operator} is defined as follows:
\begin{equation*}
  \begin{array}{l}
    \begin{array}{rcl}
      \Split (\absstates) & = & \m{split}(P,\absstates) \\
      \m{split} (\emptyset, \absstates) & = & \absstates \\
      \m{split} (\set{p} \union P', \absstates) & = &
      \letin{\cube_p = [p \mapsto 1] \kw{ and } \cube_{\lnot p} = [p \mapsto 0]}
    \end{array}\\
    \qqquad \bigcup_{\absstate \in \absstates} \m{split} (P', \pset{\absstate \sqcap \set{\cube_{\lnot p}} \sqcup
        \set{\cube}}{\cube \in_c (\absstate \sqcap
        \set{\cube_p})})\enspace.
  \end{array}
\end{equation*}
The splitting operator takes a set of Boolean heaps $\absstates$ as
arguments.  For each singleton predicate $p$ and Boolean heap
$\absstate$ it splits $\absstate$ into a set of Boolean heaps. Each of
the resulting Boolean heaps corresponds to $\absstate$, but contains
only one of the complete cubes in $\absstate$ that have a positive
occurrence of $p$. The splitting operator is sound, i.e. satisfies:
\[\gamma(\Split(P,\absstates)) \equiv \gamma(\absstates)\enspace.\] 

\smartparagraph{Cleaning.} Splitting might introduce unsatisfiable
Boolean heaps, because it is done propositionally without taking into
account the semantics of predicates. Unsatisfiable Boolean heaps
potentially lead to spurious counterexamples in the analysis and hence
should be eliminated. The same applies to cubes that are unsatisfiable
with respect to other cubes within one Boolean heap. We use a
\emph{cleaning operator} to eliminate unsatisfiable Boolean heaps and
unsatisfiable cubes within satisfiable Boolean heaps. At the same time
we strengthen the Boolean heaps with the guard of the commands before
the actual computation of the Cartesian post.  The cleaning operator
is defined as follows:
\begin{eqnarray*}
  \Clean (F, \absstates) & = &
  \letin{\absstates_1 = \pset{\absstate \in \absstates}{F \conj \gamma (\absstate) \not \models \falsum}}\\
  && \pset{\pset{\cube \in_c \absstate}{F
    \land \gamma (\absstate) \land \gamma (\cube) \not \models \falsum}}{\absstate \in \absstates_1} \enspace.
\end{eqnarray*}
The operator $\Clean$ takes as arguments a formula $F$ (e.g. the guard
of a command) and a set of Boolean heaps. It first removes all Boolean
heaps that are unsatisfiable with respect to $F$. After that it
removes from each remaining Boolean heap $\absstate$ all complete
cubes which are unsatisfiable with respect to $F$ and $\absstate$. The
cleaning operator is sound, i.e. strengthens $\absstates$ with respect
to $F$:
\[F \land \gamma(\absstates) \models \gamma(\Clean(F,\absstates))
\models \gamma(\absstates) \enspace.\]

\smartparagraph{Abstract post operator.} Figure \ref{fig:AbstractPost}
defines the abstract post operator used in Bohne. It is defined as the
composition of the splitting, cleaning, and the Cartesian post
operator.  The function $\kappa$ is a \emph{context operator}. A context
operator is a monotone mapping from sets of Boolean heaps to a context
formula. It controls the trade-off between precision and efficiency of
the abstract post operator. Our choice of $\kappa$ is described in the next
section. Figure \ref{fig:AbstractPost} also defines the abstraction
function that is used to compute the initial set of Boolean heaps. For
abstracting a formula $F$ the function $\m{abstract}$ first computes a
Boolean heap $\absstate$ which is the complement of an
under-approximation of $\lnot F$. It then splits $\absstate$ with respect
to singleton predicates and strengthens the result by the original
formula $F$. We compute the abstraction indirectly because it allows
us to reuse all the functionality that we need for computing the
abstract post operator.  We also avoid computing the best abstraction
function for the abstract domain, because the computational overhead
is not justified in terms of the gained precision.

%
%
\begin{figure}
  \begin{equation*}
    \begin{array}{l}
    \begin{array}{rcl}
      \m{abstract}(F) & = & \letin{\absstate = \compl{\pset{\cube}{\cube \models \lnot
          F}}} \\
    && \Clean (F, \Split (\absstate))
    \end{array}
    \\\\
    \kw{proc}\ \m{AbstractPost}(
    \begin{array}[t]{@{}l}
      c : \mbox{guarded command},\\
      \m{context} : \AbsDom,\\
      \absstates_0 : \AbsDom) : \AbsDom =
    \end{array}\\
    \quad
    \begin{array}{l}
      \kw{let}\ \absstates = \Clean (\guard(c), \Split (\absstates_0))\\
      \kw{let}\ \context =  \kappa (\m{context} \sqcup \absstates)\\
      \kw{return}\ \m{CartesianPost}(c, \context, \absstates)\\
    \end{array}
    \end{array}
    \end{equation*}
  \caption{Bohne's abstract post operator}
  \label{fig:AbstractPost}
\end{figure}

Assuming that $\kappa$ is in fact a context operator, soundness of
$\m{AbstractPost}$ follows from the soundness of all its component
operators. Note that soundness is still guaranteed if the underlying
validity checker is incomplete.


\section{Quantifier Instantiation}  

The context information used to strengthen the abstraction is given by
the set of Boolean heaps that are already discovered at the respective
program location. If we take into account all available context for
the abstraction of a transition then we need to recompute the abstract
transition relation in every iteration of the fixed point computation.
Otherwise the analysis would be unsound. In order to avoid unnecessary
recomputations we use the operator $\kappa$ to abstract the context by a
context formula that less likely changes from one iteration to the
next. For this purpose we introduce a domain-specific quantifier
instantiation technique. We use this technique not only in connection
with the context operator, but more generally to eliminate any
universal quantifier in a decision procedure query that originates
from the concretization of a Boolean heap. This eliminates the need
for the underlying decision procedures to deal with quantifiers.

We observed that the most valuable part of the context is the
information available over objects pointed to by program variables.
This is due to the fact that transitions always change the heap with
respect to these objects. We therefore instantiate Boolean heaps to
objects pointed to by stack variables. Bohne automatically adds an
abstraction predicate of the form $(x = v)$ for every object-valued
program variable $x$. A syntactic backwards analysis of the
procedure's postconditions is used to determine which of these
predicates are relevant at each program point.

Figure~\ref{fig:QuantInst} defines the function $\m{instantiate}$ that
uses the above mentioned predicates to instantiate a Boolean heap
$\absstate$ to a quantifier free formula (assuming abstraction
predicates itself are quantifier free). For every program variable $x$
it computes the least upper bound of all cubes in $\absstate$ which
have a positive occurrence of predicate $(x = v)$. The resulting cube
is concretized and the free variable $v$ is substituted by program
variable $x$. The function $\kappa$ maps a set of Boolean heaps
$\absstates$ to a formula by taking the join of $\absstates$ and
instantiating the resulting Boolean heap as described above. One can
shown that $\kappa$ is indeed a context operator, i.e. $\kappa$ is monotone and
the resulting formula is a context formula for $\absstates$.

\begin{figure}
  \begin{align*}
      & \Var - \mbox{object-valued program variables} \\
      & \m{instantiate}(\absstate : \mbox{Boolean heap})
       : \mbox{formula} = \\
      & \quad \letin{\m{cube}(x) = \bigsqcup (H \sqcap \set{[(x=v) \mapsto 1]})} \\
      & \ \Conj_{x \in \Var} \gamma(\m{cube}(x))[v := x]\\
      & \kappa (\absstates) = \letin{\absstate = \bigsqcup \absstates} \m{instantiate}(\absstate)
  \end{align*}
  \caption{\label{fig:QuantInst} Quantifier instantiation and the
    context operator $\kappa$}
\end{figure}

\section{Semantic Caching} 

Abstracting context does not avoid that abstract transition relations
have to be recomputed occasionally in later fixpoint iterations.
Whenever we recompute abstract transition relations we would like to
reuse the results from previous abstractions. We do this on the level
of decision procedure calls by caching the queries and the results of
the calls. The problem is that the context formulae are passed to the
decision procedure as part of the queries, so a simple syntactic
caching of formulas is ineffective. However, the context consists of
all discovered abstract states at the current iteration. Therefore it
changes monotonically from one iteration to the next. The monotonicity
of the context operator $\kappa$ guarantees that context formulae, too,
increase monotonically with respect to the entailment order. We
therefore cache formulas by keeping track of the partial order on the
context. Since context formulae occur in the antecedents of the
queries, this allows us to reuse negative results of entailment
checks from previous fixpoint iterations. This method is effective
because in practice the number of entailments which are invalid
exceeds the number of valid ones.

Furthermore, formulas are cached up to alpha equivalence. Since the
cache is self-contained, this enables caching results of decision
procedure calls not only across different fixpoint iterations in the
analysis of one procedure, but even across the analysis of different
procedures. This yields substantial improvements for procedures that
exhibit some similarity, which opens up the possibility of using our
analysis in an interactive context.

\section{Propagation of Precondition Conjuncts}

It often happens that parts of loop invariants literally come from the
procedure's preconditions. A common situation where this occurs is
that a procedure executes a loop to traverse a data structure
performing only updates on stack variables and after termination of
the loop the data structure is manipulated. In such a case the data
structure invariants are trivially preserved while executing the loop.
Using an expansive symbolic shape analysis to infer such invariants is
inappropriate. We therefore developed a fast but effective analysis
that propagates conjuncts from the precondition across the
procedure's control-flow graph. This propagation precedes the symbolic
shape analysis, such that the latter is able to assume the previously
inferred invariants.

The propagation analysis works as follows: it first splits the
procedure's precondition into a conjunction of formulas and assumes
all conjuncts at all program locations. It then recursively removes a
conjunct $F$ at program locations that have an incoming control flow
edge from some location where either (1) $F$ has been previously
removed or (2) where $F$ is not preserved under post of the associated
command. After termination of the analysis (none of the rules for
removal applies anymore) the remaining conjuncts are guaranteed to be
invariants at the corresponding program points. 

The preservation of conjuncts is checked by discharging a verification
condition (via decision procedure calls).  
The use of decision procedures makes this analysis more general
than the syntactic approach for computing frame conditions for loops
used in ESC/Java-like desugaring of loops
\cite{FlanaganSaxe01AvoidingExponentialExplosion}. In
particular, the propagation is still applicable in the presence of
heap manipulations that preserve the invariants in each loop-free code
fragment.

\chapter{Experiments}

We applied Bohne to verify operations on various data structures. Our
experiments cover data structures such as singly-linked lists,
doubly-linked lists, two-level skip lists, trees, trees with parent
pointers, sorted lists, and arrays. The verified properties include:
(1) simple safety properties, such as absence of null pointer
dereferences and array bounds checks; (2) complex data structure
consistency properties, such as preservation of the tree structure,
array invariants, as well as sortedness; and (3) procedure contracts,
stating e.g. how the set of elements stored in a data structure is
affected by the procedure.

Figure \ref{fig:Experiments} shows the results for a collection of
benchmarks running on a 2 GHz Pentium M with 1 GB memory.  The Jahob
system is implemented in Objective Caml and compiled to native code.
Running times include inference of loop invariants.  This time
dominates the time for a final check (using verification-condition
generator) that the resulting loop invariants are sufficient to prove
the postcondition. The benchmarks can be found on the Jahob project
web page \cite{Kuncak06JahobProjectWebPage}.

\begin{figure}
  \begin{center}
    \footnotesize
    \begin{tabular}{|l|l|r|r|r|}
      \hline
      benchmark & used DP & \ \# predicates & \# DP calls
      & running time \\
      & & total (user provided) & total (cache hits) & total (DP) \\
      \hline
      \hline
      List.reverse & MONA & 7 (2) & 371 (22\%) & 4s (72\%) \\
      \hline
      DLL.addLast & MONA & 7 (1) & 156 (13\%) & 3s (65\%) \\  
      \hline
      Skiplist.add & MONA & 16 (3) & 770 (20\%) & 35s (74\%) \\
      \hline
      Tree.add & MONA & 11 (3) & 983 (27\%) & 81s (91\%) \\
      \hline
      ParentTree.add & MONA & 11 (3) & 979 (27\%) & 83s (89\%) \\ 
      \hline
      SortedList.add & MONA, CVC lite & 11 (3) & 541 (17\%) & 18s (66\%)\\
      \hline
      Linear.arrayInv & CVC lite & 7 (5) & 882 (52\%) & 57s (97\%)\\
      \hline
    \end{tabular}
  \end{center}
  \caption{Results of Experiments}
  \label{fig:Experiments}
\end{figure}

We also examined the impact of our quantifier instantiation and
caching on the running time of the analysis. We have found that
disabling caching slows down the analysis by 1.3 to 1.5 times, while
disabling instantiation slows down the analysis by 1.2 to 3.6 times.

Note that our implementation of the algorithm is not highly tuned in
terms of aspects orthogonal to Bohne's algorithm, such as type
inference of internally manipulated Isabelle formulas.  We expect that
the running times would be notably improved using more efficient
implementation of Hindley-Milner type reconstruction.  In previous
benchmarks without type reconstruction in average 97\% of the time was
spent in the decision procedures.  The most promising directions for improving the
analysis performance are therefore 1)
deploying more efficient
decision procedures, and 2) further reducing the number of decision
procedure calls.

In addition to the presented examples, we have used the verification
condition generator to verify examples such as array-based
implementations of containers. The Bohne algorithm could also infer
loop invariants in such examples given the appropriate abstraction
predicates.

\chapter{Conclusions}

We have presented Bohne, a data structure analysis algorithm
based on symbolic shape analysis that generalizes predicate
abstraction and infers Boolean algebra expressions over sets
given by predicates on objects.  We have shown that this
abstraction can be fruitfully combined with a collection of
decision procedures that operate on
independent subgoals of the same proof obligation.  The
effect of such an approach is that the analysis synthesizes
facts that are used to communicate information between
different decision procedures.  As a result, we were able to
combine precise reasoning about reachability in tree-like
structures with reasoning about first-order properties in
general graphs and integer arithmetic properties.  
As an example
that illustrates this combination, we have verified
a sorted linked data structure without specializing
the analysis to sorting or reachability properties.

In addition, we have deployed a range of techniques that
significantly improve the running time of the analysis and
the level of automation compared to direct application of
the algorithm.  These techniques
include context-dependent finite-state abstraction,
semantic caching of formulas, propagation of conjuncts, and
domain-specific quantifier instantiation.  Our current
experience with the Bohne analysis in the context
of the Hob and Jahob data structure verification systems
suggests that it is effective for verifying a wide range of
data structures and that its running time makes it usable
for verification of such complex properties.

\bibliographystyle{abbrv}
\bibliography{pnew}

\end{document}